\documentclass[preprint,prb]{revtex4-2}

\usepackage{graphicx}
\usepackage{dcolumn}
\usepackage{float}
\usepackage{amsmath}

\begin{document}

\title{\boldmath Planckian behavior of cuprates at the pseudogap critical point simulated via flat electron-boson spectral density \unboldmath}

\author{Hwiwoo Park} \author{Jun H. Park} \author{Jungseek Hwang}
\email{jungseek@skku.edu}
\affiliation{$^1$Department of Physics, Sungkyunkwan University, Suwon, Gyeonggi-do 16419, Republic of Korea \\ $^2$School of Mechanical Engineering, Sungkyunkwan University, Suwon, Gyeonggi-do 16419, Republic of Korea}
\date{\today}

\begin{abstract}

Planckian behavior has been recently observed in La$_{1.76}$Sr$_{0.24}$CuO$_4$ at the pseudogap critical point. The Planckian behavior takes place in an intriguing quantum metallic state at a quantum critical point. Here, the Planckian behavior was simulated with an energy-independent (or flat) and weakly temperature-dependent electron-boson spectral density (EBSD) function by using a generalized Allen's (Shulga's) formula. We obtained various optical quantities from the flat EBSD function, such as the optical scattering rate, the optical effective mass, and the optical conductivity. These quantities are well fitted with the recently observed Planckian behavior. Fermi-liquid behavior was also simulated with an energy-linear and temperature-independent EBSD function. The EBSD functions agree well with the overall doping- and temperature-dependent trends of the EBSD function obtained from the optically measured spectra of cuprate systems, which can be crucial for understanding the microscopic electron-pairing mechanism for high-$T_c$ superconductivity in cuprates.
\\

\noindent {\bf Keywords:} High-$T_c$ cuprates, Planckian behavior, Electron-boson spectral density function, Allen's formula, Fermi-liquid behavior
\\

\noindent *Correspondence author \\
  Email: jungseek@skku.edu (Jungseek Hwang)

\end{abstract}

%\pacs{74.25.-q, 74.25.Gz, 74.25.Kc}

\maketitle

\section*{Introduction}

Copper oxide superconductors (cuprates) have been intensively studied since their discovery \cite{bednorz:1986,wu:1987,maeda:1988}. However, the microscopic pairing mechanism has not been found yet. The phase diagram of the cuprates has become more complicated as the research goes on \cite{keimer:2015}. However, the generic phase diagram of cuprates is well known \cite{batlogg:2000,taillefer:2010}. The undoped parent copper oxides are antiferromagnetic (AFM) insulators. As the doping increases, the AFM phase is suppressed, and the superconducting state appears. At the optimal doping ($p_{\mathrm{opt}}$), the superconducting (SC) critical temperature ($T_c$) reaches its maximum value, and further doping decreases the SC $T_c$, resulting in the $d$-wave SC dome in the phase diagram. The doping levels lower (higher) than the $p_{\mathrm{opt}}$ are called underdoped (overdoped) in the phase diagram. There is the peculiar pseudogap phase, which shows a partial suppression of the density of states near the Fermi level and is not completely understood yet. The pseudogap phase is suppressed as the doping increases and, eventually, disappears at a certain doping level on the overdoped side of the phase diagram, which is called the pseudogap critical point ($p^*$) \cite{taillefer:2010}. Recently, an optical study on La$_{2-x}$Sr$_x$CuO$_4$ (LSCO at $x$ = 0.24) at the $p^*$ demonstrated that the LSCO at $p^*$ showed the Planckian behavior \cite{michon:2023}: the normalized optical scattering rate and effective mass as functions of the normalized energy with respect to the temperature ($\hbar\omega/k_B T$) fall into universal curves. The Planckian behavior is directly associated with the $T$-linear behavior in the transport resistivity and the $\omega$-linear behavior in the optical scattering rate. The Planckian metal was proposed by Patel {\it et al.} \cite{patel:2019}. In the Planckian metal, the transport scattering rate is proportional to the Planckian frequency ($k_BT/\hbar$).  

The parent copper oxides are charge-transfer insulators. The charge-transfer gap is formed between the oxygen $p$ orbital and the upper Hubbard band, which appears due to the strong Coulomb repulsion between charge carriers. The charge carriers in the doped copper oxides may interact with each other by exchanging force-mediated bosons, which are most likely spin-1 antiferromagnetic fluctuations based on the generic phase diagram of cuprates. The interaction spectrum between charge carriers can be obtained from the measured spectrum via various spectroscopic techniques such as infrared spectroscopy, angle-resolved photoemission spectroscopy, tunneling spectroscopy, scanning tunneling microscopy, Raman spectroscopy, and inelastic neutron spectroscopy \cite{carbotte:2011}. The interaction spectrum is called the glue function because it is expected to carry information about the bonding force for the Cooper-pair formation in the cuprates. The interaction spectrum is also known as the electron-boson spectral density (EBSD) function. Among the spectroscopic techniques, infrared spectroscopy has crucial advantages because it can be used for studying all cuprate systems with high energy resolution and, importantly, provides absolute quantitative results. The doping- and temperature-dependent EBSD functions have been obtained from the measured infrared spectra of cuprates \cite{carbotte:1999,schachinger:2000,hwang:2006,hwang:2007,heumen:2009,hwang:2021}. The EBSD function of cuprates shows characteristic systematic doping- and temperature-dependent evolutions (see Fig. \ref{fig5}). In the underdoped cuprates, the most spectral weight of the EBSD function is confined to the low energy region at low temperatures. As the doping increases, the spectral weight spreads over a wide spectral range, resulting in a less energy-dependent EBSD function. The highly overdoped cuprate shows a nearly energy-independent (or flat) EBSD function \cite{hwang:2007,hwang:2021} (also see Fig. \ref{fig5}). As the temperature increases, the spectral weight in the low energy region transfers to the high-frequency region, resulting in less frequency dependence. As the doping increases, the intensity of the temperature-dependent spectral redistribution gets weaker. The highly overdoped cuprate shows a very small temperature-dependent EBSD function. Therefore, the EBSD function of highly overdoped cuprate is nearly energy-independent (or flat) and weakly temperature-dependent. Note that the discussions on the EBSD function are for the normal state. Planckian behavior was observed in highly overdoped LSCO cuprate near the pseudogap critical point ($p^*$) \cite{michon:2023} (see also Fig. \ref{fig5}).

In this study, a nearly energy-independent and weakly temperature-dependent EBSD function was used to simulate Planckian behavior. A proposed reverse process \cite{hwang:2015a} was used to obtain the optical quantities from the model EBSD function as follows: From the model EBSD function, the optical scattering rate was obtained using a generalized Allen's (or Shulga's) formula \cite{shulga:1991}. The optical scattering rate is the imaginary part of the optical self-energy \cite{hwang:2004}. The real part of the optical self-energy was obtained from the imaginary part using the Kramers-Kronig relation between them. Then, the complex optical conductivity was obtained from the complex optical self-energy using the extended Drude model \cite{gotze:1972,allen:1977,puchkov:1996,hwang:2004}. We found that the obtained optical self-energy and optical conductivity clearly showed Planckian behavior. We also simulated Fermi-liquid behavior with an energy-linear and temperature-independent EBSD function. The EBSD functions are well fitted to the doping- and temperature-dependent evolutions of the EBSD function obtained from the optically measured spectra of cuprate systems.

\section*{Results and discussion}

\subsection*{Model EBSD function and optical quantities}

\begin{figure}[!htbp]
  \vspace*{-0.1 cm}%
 \centerline{\includegraphics[width=5.5 in]{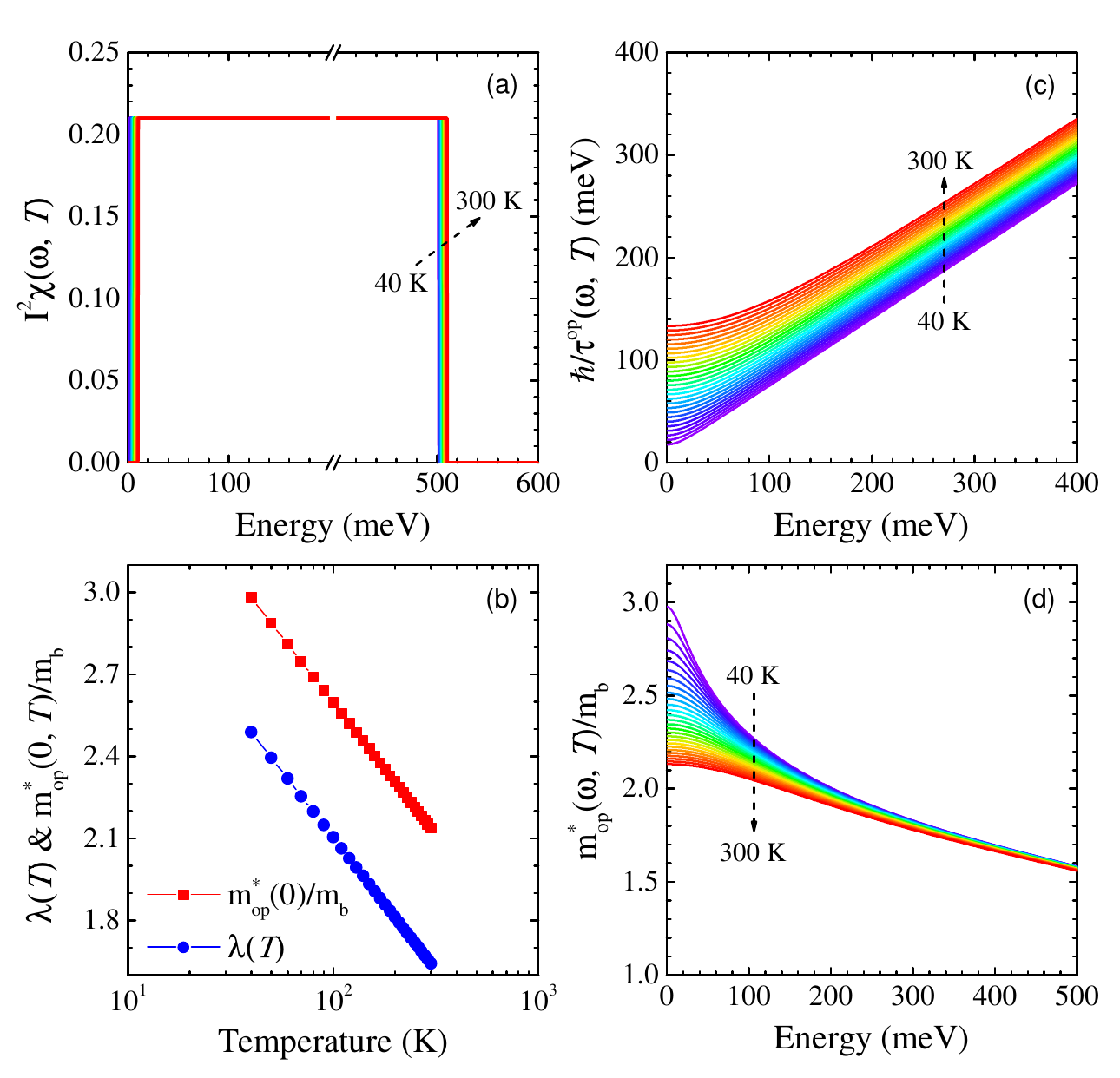}}%
  \vspace*{-0.3 cm}%
\caption{(Color online) (a) The model electron-boson spectral density (EBSD) function at various temperature from 40 to 300 K with a 10 K increment. (b) The coupling constant ($\lambda(T)$) and effective mass ($m^*_{op}(0, T)/m_b$) as functions of temperature. (c) The optical scattering rate (or the imaginary part of the optical self-energy) at various temperatures from 40 to 300 K with a 10 K increment. (d) The corresponding optical effective mass ($m^*_{op}(\omega, T)/m_b$) at various temperatures. }
\label{fig1}
\end{figure}

We start from an energy-independent (or flat) and weakly temperature-dependent model EBSD function. The model EBSD function ($I^2\chi(\omega,T)$) can be written as:
\begin{eqnarray}
I^2\chi(\omega, T) &=& A \:\:\:\mbox{for} \:\:\: \omega_L(T) \leq \omega \leq \omega_H+\omega_L(T) \\ \nonumber
 &=& 0 \:\:\: \mbox{otherwise},
\end{eqnarray}
where $I$ is the coupling parameter between the mediated boson and charge carrier, $\chi(\omega, T)$ is the bosonic spectrum, $A$ is a constant, $\omega_H$ is a constant frequency, and $\omega_L(T)$ is a temperature-dependent frequency. The coupling constant ($\lambda(T)$) is defined as $\lambda(T) \equiv 2\int_0^{\omega_c} [I^2\chi(\Omega,T)/\Omega] d\Omega$, where $\omega_c$ is the cutoff frequency, in our case $\omega_c(T) =  \omega_L(T) + \omega_H$. Requiring that $\lambda(T)$ is linear to $-\ln T$, we determine the $\omega_L(T)$ as $\omega_L(T) = \omega_H (C/T-1)^{-1}$, where $C$ is a constant with a unit of kelvin (K) (refer to the Supplementary Materials for a more detailed description of the determination of $\omega_L(T)$). The coupling constant can be analytically calculated as $\lambda(T) = 2A(\ln{C}-\ln{T})$. Note that the coupling constant is proportional to $-\ln{T}$ with the proportionality constant $2A$, an adjustable parameter. Fig. \ref{fig1}(a) shows the model $I^2\chi(\omega, T)$ with $\omega_H =$ 500 meV, $A$ = 0.21, and $C$ = 15 eV  at various temperatures from 40 to 300 K with a 10 K increment. In Fig. \ref{fig1}(b), the coupling constant ($\lambda(T)$) is shown, i.e. $\lambda(T) \cong 5.068-0.42 \ln{T} = 5.068 - 0.96 \log{T}$.

The optical scattering rate ($\hbar/\tau^{op}(\omega, T)$) is related to the EBSD function ($I^2\chi(\omega, T)$) through the generalized Allen's (or Shulga's) formula\cite{shulga:1991} as:
\begin{equation}
\frac{\hbar}{\tau^{op}(\omega, T)} = \hbar \int_0^{\infty} I^2\chi(\Omega,T) K(\omega, \Omega, T)d\Omega,
\end{equation}
where $K(\omega, \Omega, T)$ is the kernel, which depends on the material state \cite{hwang:2015a}. In this study, because we deal with the normal state with a constant density of states \cite{shulga:1991}, $K(\omega, \Omega, T) = \frac{\pi}{\omega} \Big{\{} 2 \omega \coth{\Big{[}\frac{\hbar \Omega}{2k_B T}\Big{]}} - (\omega+\Omega) \coth{\Big{[}\frac{\hbar (\omega+\Omega)}{2k_B T}\Big{]}} + (\omega-\Omega) \coth{\Big{[}\frac{\hbar (\omega-\Omega)}{2 k_B T}\Big{]}} \Big{\}}$, where $\hbar$ is the reduced Planck constant and $k_B$ is the Boltzmann constant. Fig. \ref{fig1}(c) shows the optical scattering rates at various temperatures obtained from the model $I^2\chi(\omega, T)$ using the Shulga's formula (Eq. (2)). As the temperature increases, $\hbar/\tau^{op}(\omega, T)$ monotonically increases and shows an energy-linear dependence in the high energy region up to $\sim$500 meV. The scattering rates in a quite wide spectral range up to 1000 eV are shown in the Supplementary Materials (Fig. S1(a)).

The optical scattering rate ($\hbar/\tau^{op}(\omega, T)$) corresponds to the imaginary part of the complex optical self-energy ($-2\tilde{\Sigma}^{op}(\omega, T) \equiv [-2\Sigma^{op}_1(\omega, T)] + i[-2\Sigma^{op}_2(\omega, T)]$), which is defined by the extended Drude model \cite{gotze:1972,allen:1977,puchkov:1996,hwang:2004} as:
\begin{eqnarray}
  \tilde{\sigma}(\omega, T) &=& i\frac{\hbar \Omega_p^2}{4 \pi}\frac{1}{\hbar \omega + [-2\tilde{\Sigma}^{op}(\omega, T)]}, \\ \nonumber
  &=& i\frac{\hbar \Omega_p^2}{4 \pi} \frac{1}{\omega [1 + \lambda^{op}(\omega, T)] + i 1/\tau^{op}(\omega, T)},
\end{eqnarray}
where $\tilde{\sigma}(\omega, T)$ $(\equiv \sigma_1(\omega, T) + i\sigma_2(\omega, T))$ is the complex optical conductivity, $\Omega_p$ is the plasma frequency of charge carriers, and $\lambda^{op}(\omega, T)$ is the optical coupling function. The optical coupling function is related to the optical effective mass ($m^*_{op}(\omega, T)$) as $\lambda^{op}(\omega, T) + 1 \equiv m^*_{op}(\omega, T)/m_b$, where $m_b$ is the band mass. The real and imaginary parts of the complex optical self-energy form a Kramers-Kronig pair \cite{hwang:2004,hwang:2015a}. Therefore, the real part can be obtained from the imaginary part using the Kramers-Kroing relation between them, i.e., $-2\Sigma^{op}_1(\omega, T) = -(2/\pi)P\int_0^{\infty}[-2\Sigma^{op}_2(\omega', T)]/[\hbar(\omega'^2-\omega^2)]d\omega'$, where $P$ represents the principal part of the improper integral. The obtained $-2\Sigma^{op}_1(\omega, T)$ in a wide spectral range at various temperatures are shown in the Supplementary Materials (Fig. S1(b)). The obtained optical effective masses ($m^*_{op}(\omega, T)/m_b \equiv [-2\Sigma^{op}_1(\omega, T)]/\hbar\omega + 1$) at various temperatures are shown in Fig. \ref{fig1}(d). As the temperature decreases, the effective mass increases and shows a strong temperature dependence in the low energy region. The optical effective masses in a wide spectral range are shown in the Supplementary Materials (Fig. S1(c)). The optical effective masses at zero ($m^*_{op}(0, T)/m_b$) are shown in Fig. \ref{fig1}(b), which shows a $-\ln{T}$ dependence as in the published literature \cite{michon:2023}.

\begin{figure}[!htbp]
  \vspace*{-0.1 cm}%
 \centerline{\includegraphics[width=5.5 in]{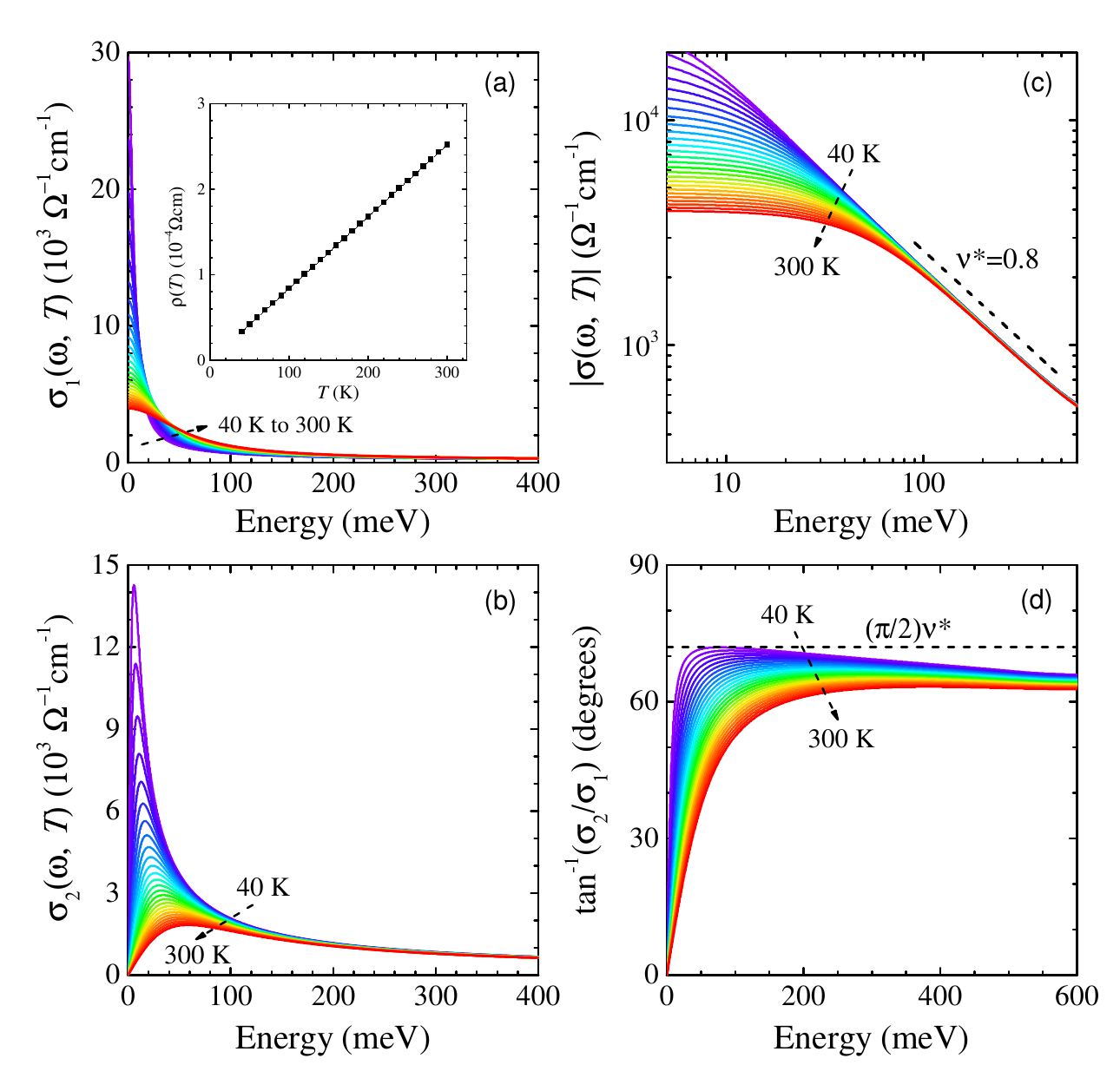}}%
  \vspace*{-0.3 cm}%
\caption{(Color online) The complex optical conductivity. (a) and (b) show the real and imaginary parts of the optical conductivity at various temperatures, respectively. (c) and (d) show the amplitude and phase of the complex optical conductivity at various temperatures, respectively.}
\label{fig2}
\end{figure}

Further, the complex optical conductivity was obtained from the complex optical self-energy using the extended Drude model (Eq. (3)) with a plasma frequency of 2 eV. The real ($\sigma_1(\omega, T)$) and imaginary ($\sigma_2(\omega, T)$) parts of the complex optical conductivity at various temperatures are shown in Fig. \ref{fig2}(a) and (b), respectively. The DC resistivity ($\rho(T)$) was obtained by taking an extrapolation of the real part of the optical conductivity to zero, i.e., $\rho(T) \equiv 1/\sigma_1(0, T)$. The inset of Fig. \ref{fig2}(a) shows the DC resistivity ($\rho(T)$) as a function of temperature, which exhibits a $T$-linear behavior. The amplitude ($|\tilde{\sigma}(\omega, T)| = \sqrt{[\sigma_1(\omega, T)]^2+[\sigma_2(\omega, T)]^2}$) and phase ($\mbox{arg}(\tilde{\sigma}(\omega, T))=\tan^{-1}[\sigma_2(\omega, T)/\sigma_1(\omega, T)]$) of the complex optical conductivity were obtained. The amplitude and phase of the optical conductivity are shown in Figs. \ref{fig2}(c) and \ref{fig2}(d), respectively. The $\tilde{\sigma}(\omega)$ shows $D(-i\omega)^{-\nu^*} = D\omega^{-\nu^*}e^{i(\pi/2)\nu^*}$ behavior \cite{marel:2003} with an exponent $\nu^* =$ 0.8, where $D$ is a constant. The $\nu^*$ was obtained from the absolute value of the slope in the amplitude between 100 and 500 meV. This value is the same as that in the published literature \cite{michon:2023}. The amplitude and phase in a wide spectral range are shown in the Supplementary Materials (Figs. S2(a) and S2(b)).

\subsection*{Planckian behavior}

Now, we check whether the optical quantities obtained from the model EBSD function exhibit Planckian behavior. The Planckian scaling behaviors of $\hbar/\tau^{op}(\omega, T)$ and $m^*_{op}(\omega, T)/m_b$ can be described as follows \cite{michon:2023}:
\begin{eqnarray}
  \frac{\hbar}{\tau^{op}(\omega, T)} &=& (k_B T)^{\nu}f_{\tau}\Big{(} \frac{\hbar \omega}{k_B T} \Big{)}, \\ \nonumber \frac{m^*_{op}(\omega, T)}{m_b} - \frac{m^*_{op}(0, T)}{m_b} &=& (k_B T)^{\nu - 1}f_m\Big{(} \frac{\hbar \omega}{k_B T} \Big{)},
\end{eqnarray}
where $f_{\tau}(x)$ and $f_m(x)$ are two scaling functions for the scattering rate and the effective mass, respectively. The equations above are valid for $\nu \leq 1$ \cite{berthod:2013,michon:2023}. When $\nu =$ 1, the scaling behavior is called the Planckian one. When $\nu < 1$, the scaling behavior is called the sub-Planckian behavior \cite{michon:2023}. If $\nu >1$ the scaling for the scattering rate holds but that of the effective mass does not hold (see Figs. \ref{fig4}(c) and \ref{fig4}(d)).

\begin{figure}[!htbp]
  \vspace*{-0.2 cm}%
 \centerline{\includegraphics[width=5.5 in]{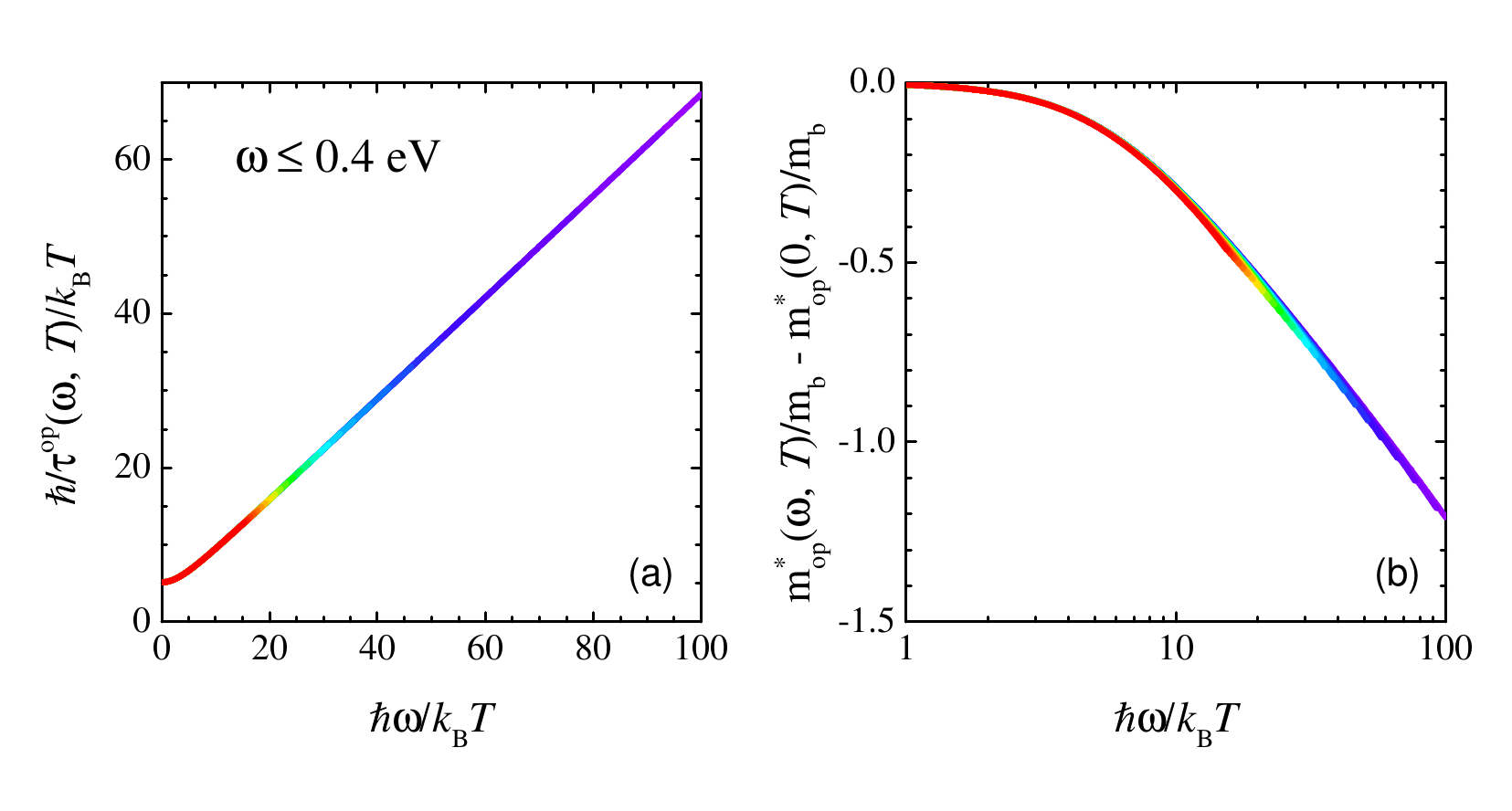}}%
  \vspace*{-0.5 cm}%
\caption{(Color online) The Planckian behavior. (a) and (b) are the scaling functions of $\hbar \omega/ (k_B T)$ for the $\hbar/\tau^{op}(\omega, T)$ and the $m^*_{op}(\omega, T)/m_b$, respectively. }
\label{fig3}
\end{figure}

$[\hbar/\tau^{op}(\omega, T)]/k_B T$ and $m^*_{op}(\omega, T)/m_b - m^*_{op}(0, T)/m_b$ as functions of $\hbar \omega/k_B T$ are shown in Figs. \ref{fig3}(a) and \ref{fig3}(b), respectively. For each quantity, all curves at different temperatures fall into nearly a single scaling curve with $\nu =$ 1; $\hbar/\tau^{op}(\omega, T)$ and $m^*_{op}(\omega, T)/m_b$ show the Planckian behavior. Therefore, the model EBSD function results in Planckian behavior. Because Planckian behavior was observed in LSCO at a highly overdoped level (or the pseudogap critical point, $p^*$) \cite{michon:2023}, the model EBSD function is expected to be the EBSD function of a highly overdoped cuprate. In this regard, the model EBSD function is well fitted in the overall doping- and temperature-dependent EBSD function obtained from the measured optical spectra of cuprates \cite{hwang:2007,hwang:2021} in terms of the intensity level, energy independence, and weak temperature dependence. The model energy-independent EBSD function may be associated with the marginal Fermi-liquid behavior \cite{varma:1989,littlewood:1991}. Norman and Chubukov found that the high-frequency behavior obtained using by a broad (or small energy-dependent) EBSD function could be suggested evidence for quantum-critical scaling \cite{norman:2006}. Because the EBSD function spreads in a wide spectral range at high temperatures such as 300 K, the measured optical spectra of cuprates in a wide doping range at 300 K have been analyzed using the marginal Fermi-liquid model \cite{hwang:2004a}. However, Planckian scaling behavior was observed in LSCO at the pseudogap critical point ($p^*$) in a wide temperature range \cite{michon:2023}. 

According to recent studies \cite{bruin:2013,legros:2019}, the $T$-linear growth of the scattering rate of electrons ($1/\tau(T)$) can be expressed as $1/\tau(T) = \alpha k_B T/\hbar$, where $\alpha$ is a universal parameter that varies weakly between materials and is roughly 0.7 - 1.1. This weak material-dependent and $T$-linear scattering rate behavior was called universal $T$-linear scattering rate (or resistivity). However, Sadovskii criticized the claim that the universal $T$-linear resistivity was a kind of delusion related to a particular procedure to present the experimental data and that the parameter $\alpha$ is inherently proportional to the strength of interaction \cite{sadovskii:2021}. Consequently, whether or not the parameter $\alpha$ is universal is an important question. Thus far in our investigation, we have proven that the Planckian scaling behavior was obtained by using the unique EBSD function depicted in Fig. \ref{fig1}(a). At $\omega = 0$, the Planckian scaling behavior of the optical scattering rate is reduced to $\hbar/\tau^{op}(0, T) \equiv \hbar/\tau(T) = f_{\tau}(0) k_B T $ from Eq. (4) with $\nu = 1$. In this instance, the parameter $\alpha$ is equal to $f_{\tau}(0)$, which is a constant of $\sim$5 (see Fig. \ref{fig3}(a)) and depends on the EBSD function, i.e., the strength of interaction. As a result, our findings suggest that the parameter $\alpha$ is dependent on interaction strength rather than being a universal constant.

One may ask about the relationship between our study and the study presented in Ref. \cite{michon:2023}. Both studies have simulated the experimentally observed optical spectra, which exhibit Planckian scaling behavior. However, the starting quantities are different between the two studies. In Ref. \cite{michon:2023}, the authors started with the imaginary part of the quasiparticle self-energy with a special shape, which obeys an $\omega/T$ scaling property. But, in our study, we started with the EBSD function with a special shape, which is more fundamental in a sense and easily compared with optical results. Therefore, we found the special-shaped EBSD function in terms of temperature and energy, which gives rise to Planckian scaling behavior. Furthermore, the EBSD function fits well with the experimental EBSD functions in terms of temperature and doping, as discussed in the following section.

\subsection*{Fermi-liquid behavior}

\begin{figure}[!htbp]
  \vspace*{-0.3 cm}%
 \centerline{\includegraphics[width=5.0 in]{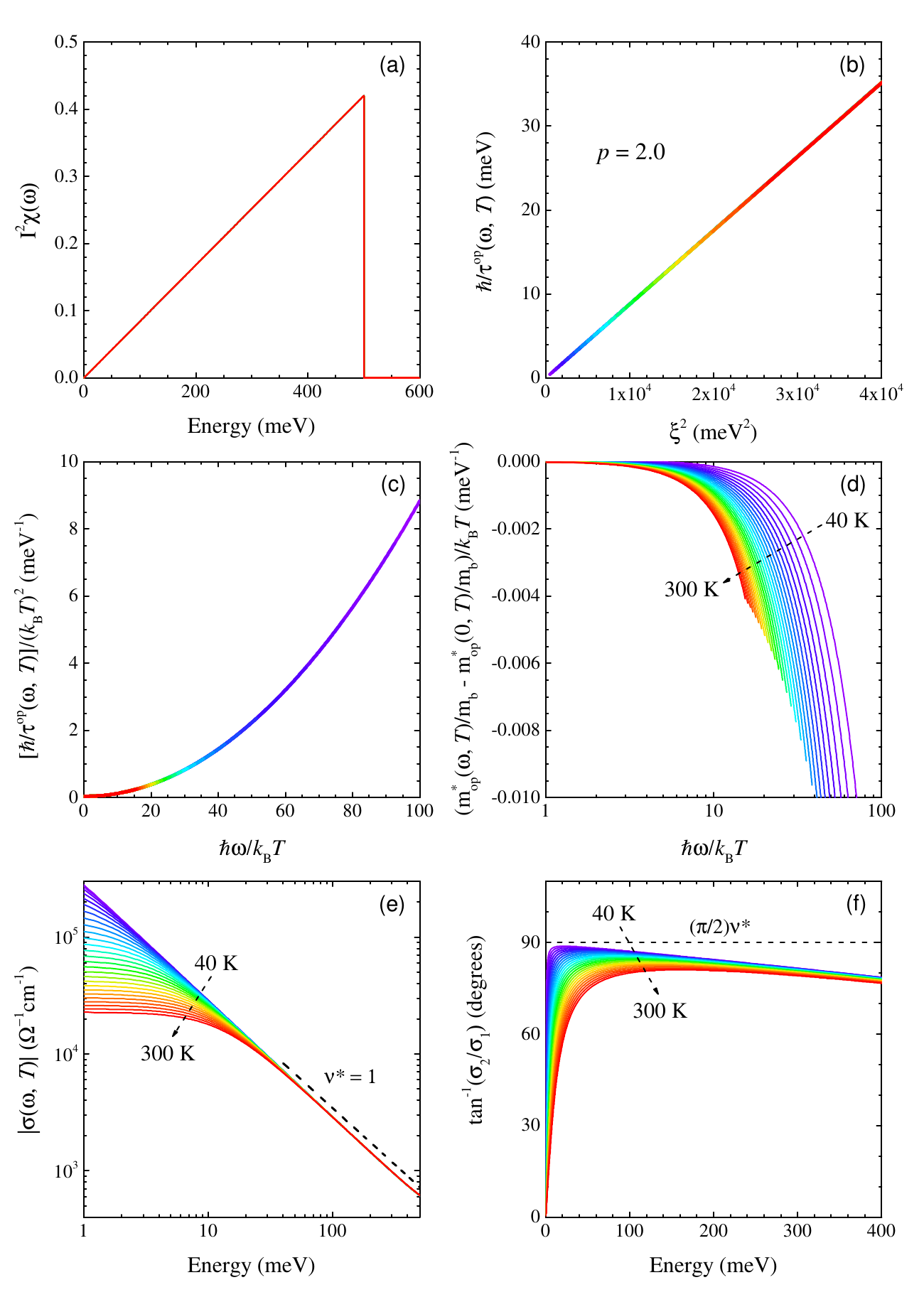}}%
  \vspace*{-0.3 cm}%
\caption{(Color online) Fermi-liquid behavior. (a) The temperature-independent EBSD function. (b) A universal curve of the optical scattering rate, i.e., $\hbar/\tau^{op}(\omega, T)$ versus $\xi^2 (= (\hbar\omega)^2+(2p\pi k_B T)^2)$ with $p =$ 2. (c) The scaling function of the optical scattering rate with $\nu =$ 2, i.e., $\hbar/\tau^{op}(\omega, T)/(k_B T)^2$ versus $\hbar \omega/k_B T$. (d) The scaling function of the effective mass with $\nu =$ 2, i.e., $[m^*_{op}(\omega, T)/m_b - m^*_{op}(0, T)/m_b]/(k_B T)^2$ versus $\hbar \omega/k_B T$.  (e) and (f) show the amplitude and phase of the complex optical conductivity, respectively.}
\label{fig4}
\end{figure}

\begin{figure}[!htbp]
  \vspace*{-0.3 cm}%
 \centerline{\includegraphics[width=5.4 in]{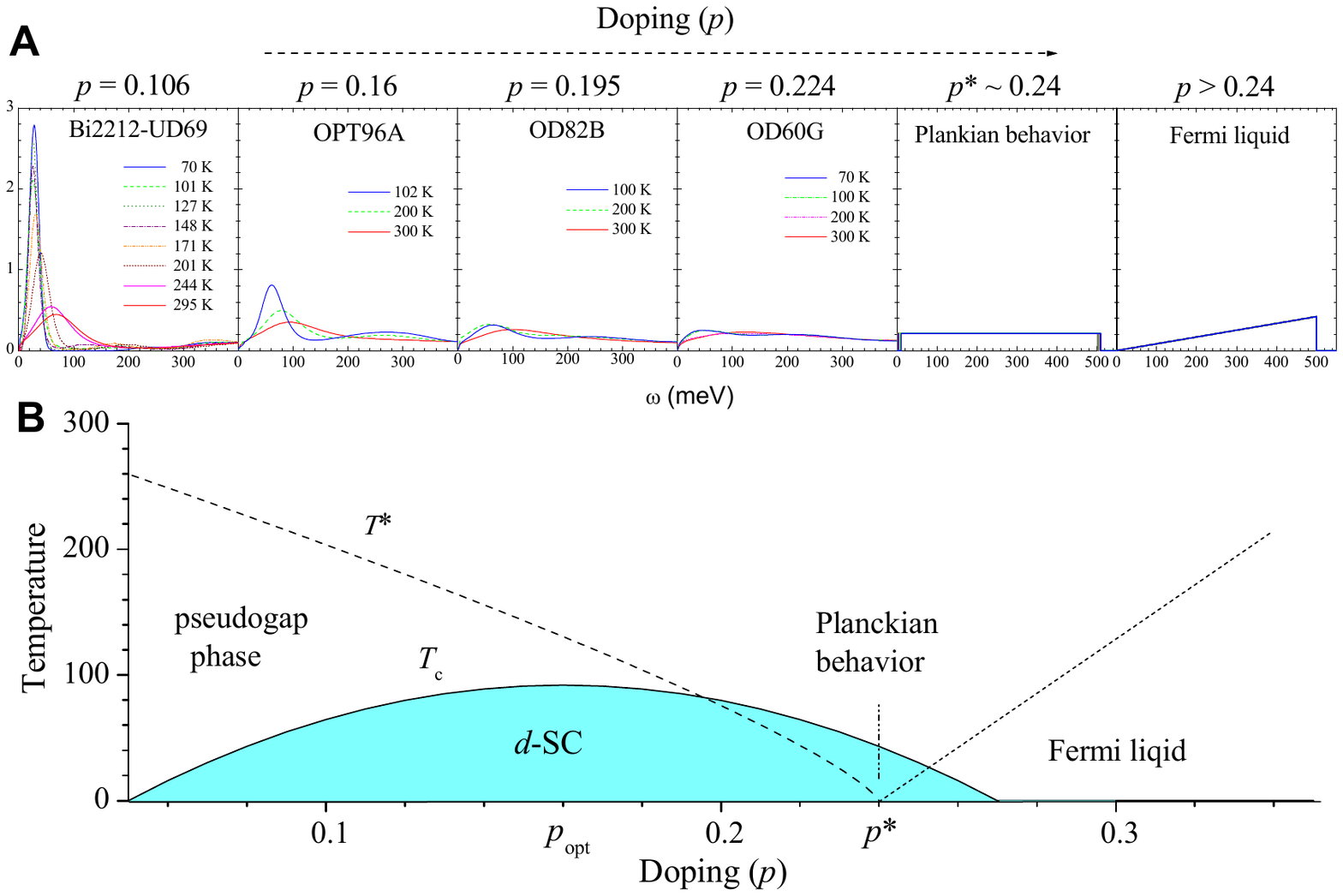}}%
  \vspace*{-0.3 cm}%
\caption{(Color online) (a) The doping- and temperature-dependent EBSD function of hole-doped Bi$_2$Sr$_2$CaCu$_2$O$_{8+\delta}$ cuprate \cite{hwang:2007,hwang:2021}, including the Planckian and Fermi-liquid behaviors. (b) A schematic doping-temperature ($p$-$T$) phase diagram of hole-doped cuprates. Note that $p_{\mathrm{opt}}$ and $p^*$ are the optimal doping level and the pseudogap critical point, respectively.}
\label{fig5}
\end{figure}

We also obtained the Fermi-liquid behavior from another model EBSD function, whose state is expected to be located in a very overdoped region outside of the $d$-wave SC dome of the phase diagram of cuprates. The new model EBSD function is as follows: $I^2\chi(\omega, T) = (C/\omega_c)\: \omega$ for $\omega \leq \omega_c$ and 0 otherwise, where $C$ is a constant and $\omega_c$ is the cutoff frequency. In our calculations, $C = 0.42$ and $\omega_c =$ 500 meV. Note that the model EBSD function is energy-linear and temperature-independent. The EBSD function is shown in Fig. \ref{fig4}(a). Note that, up to now, any optical measurements for the cuprates have not given rise to an energy-linear and temperature-independent EBSD function yet. The coupling constant ($\lambda(T)$) as a function of temperature is a constant ($\sim$0.84) as shown in the Supplementary Materials (Fig. S5(a)). The resulting optical quantities at various temperatures from 40 to 300 K with a 10 K increment are summarized in Figs. \ref{fig4}(b-f). Fig. \ref{fig4}(b) shows another scaling function for the optical scattering rate \cite{mirzaei:2013}, i.e., $\hbar/\tau^{op}(\omega, T)$ versus $\xi^2 \equiv (\hbar\omega)^2 + (p \pi k_B T)^2$ with $p$ =2, where $p$ is an adjustable parameter and the Fermi-liquid behavior is expected if $p =$ 2. Because all the optical scattering rates fall into a single curve, the resulting optical scattering rate clearly shows Fermi-liquid behavior. Fig. \ref{fig4}(c) shows the scaling function (Eq. (4)) for the optical scattering rate with $\nu = 2$, i.e., $[\hbar/\tau^{op}(\omega, T)]/(k_B T)^2$ versus $\hbar \omega/k_B T$, where all the optical scattering rates at various temperatures fall into a single curve. Fig. \ref{fig4}(d) shows the scaling function for the optical effective mass with $\nu = 2$, i.e., $[m^*_{op}(\omega, T) - m^*_{op}(0, T)]/k_B T$ versus $\hbar \omega/k_B T$ at various temperatures. In the case of the optical effective mass, the curves at various temperatures do not fall into a single curve. Note that the optical effective mass at zero ($m^*_{op}(0, T)/m_b$) as a function of temperature is shown in the Supplementary Materials (Fig. S5(a)), where the $m^*_{op}(0, T)/m_b$ is nearly temperature-independent. All the real parts of optical self-energy in the low-frequency region are proportional to the energy ($\hbar\omega$) with nearly the same proportionality constant (see Fig. S3(b)). The amplitude and phase of the complex optical conductivity are shown in Figs. \ref{fig4}(e) and \ref{fig4}(f), respectively. We observed a larger exponent $\nu^*$ than that for the Planckian behavior. The absolute value ($\nu^*$) of the slope in the amplitude between 100 and 500 meV is $1.0$. It was observed that the exponent increases as the doping increases in the literature \cite{hwang:2007a}. Therefore, the Fermi-liquid behavior is expected to be at a higher doping level than the Planckian one.

The optical scattering rate, the real part of the optical self-energy, and the optical effective mass are shown in the Supplementary Materials (Figs. S3(a-f)). The complex optical conductivity, the amplitude, and the phase in a wide spectral range are shown in the Supplementary Materials (Figs. S4(a-d)). The DC resistivity as a function of temperature square ($T^2$) obtained from the optical conductivity exhibits a $T$-quadratic dependence and is shown in the Supplementary Materials (Fig. S5(b)). We also compared the reflectance spectra at various temperatures for the Planckian and Fermi-liquid behaviors in the Supplementary Materials (Fig. S6). The reflectance spectra exhibit some characteristic energy- and temperature-dependent trends; reflectance for the Planckian behavior exhibits more temperature dependence and a roughly energy-linear dependence, whereas that for the Fermi-liquid behavior exhibits a higher reflectance, less $T$-dependence, and a roughly $\omega$-quadratic dependence. The higher reflectance of the Fermi-liquid behavior than the Planckian one indicates that the Fermi-liquid behavior is at a higher doping level than the Planckian behavior. 

One may expect that the EBSD function varying with $\omega^n$ yields the optical self-energy varying with $\omega^{n+1}$. Our study shows that the expectation is more or less true, the frequency dependencies of the model EBSD functions in a finite spectral range are important, and, additionally, their temperature dependencies are closely involved in simulating the Planckian and Fermi-liquid behaviors. Our study demonstrated that the optical scattering rates, which show the Planckian and Fermi-liquid behaviors, could be generated using suitable choices of temperature-and energy-dependent model EBSD functions. The EBSD function spectrum in cuprates has been known to be intimately associated with the pairing glue spectrum. Therefore, its detailed temperature- and doping-dependent properties can be crucial for figuring out the microscopic origin of the pairing glue, which is still elusive.

In Fig. \ref{fig5}, the doping- and temperature-dependent EBSD function of hole-doped Bi$_2$Sr$_2$CaCu$_2$O$_{8+\delta}$ cuprate is summarized by including the Planckian and Fermi-liquid behaviors and the phase diagram for showing them. It is worth noting that the EBSD functions in the figure are in their normal state. In the SC state, the sharp optical coherence mode at low frequencies is significantly enhanced \cite{hwang:2007,hwang:2021}. As we previously mentioned, the doping- and temperature-dependent evolutions can be clearly seen; as both doping and temperature increase, the spectral weight shifts from low to high energy, making the spectral weight spread through the wide spectral range, resulting in a frequency-independent and weakly temperature-dependent EBSD function. The EBSD functions for Planckian and Fermi-liquid behaviors are well fitted with the doping- and temperature-dependent trends of the experimental EBSD function based on the amplitude, frequency dependence (or shape), and temperature dependence. In the Planckian state, there is no sign of the sharp optical coherence mode in the EBSD function, at least in the normal state, which seems to be consistent with the previous optical observations \cite{hwang:2004,hwang:2007}, where the optical mode disappeared in the SC dome.

\section*{Conclusion}

We investigated the Planckian behavior that was observed in LSCO at the pseudogap critical point ($p^*$). We simulated Planckian behavior with an energy-independent (or flat) and weakly temperature-dependent EBSD function using a generalized Allen's (or Shulga's) formula. The model EBSD function for Planckian behavior is expected to be in a highly overdoped region. The Planckian scaling functions for the optical scattering rate and the optical effective mass were clearly simulated from the model EBSD function. We also proposed another model EBSD function, which is energy-linear and temperature-independent, for Fermi-liquid behavior. The Fermi-liquid behavior is expected in a very highly overdoped region above the SC dome in the phase diagram of cuprates. The results from the two different model EBSD functions were compared with one another. We found that, in some aspects, the Fermi-liquid behavior is at a higher doping level in the phase diagram of cuprates than the Planckian behavior. Therefore, the two proposed EBSD functions agree well with the overall doping- and temperature-dependent trends of the EBSD function of cuprates, which can be important for solving the conundrum of the microscopic electron-pairing mechanism for high-$T_c$ superconductivity in cuprates. 
\\

\noindent {\bf Acknowledgements} J.H. acknowledges the financial support from the National Research Foundation of Korea (NRFK Grant Nos. 2021R1A2C101109811). This research was also supported by BrainLink program funded by the Ministry of Science and ICT through the National Research Foundation of Korea (2022H1D3A3A01077468).
\\

%
% bibliography
%
\bibliographystyle{naturemag}
\bibliography{bib}% Produces the bibliography via BibTeX.

\begin{thebibliography}{10}
\expandafter\ifx\csname url\endcsname\relax
  \def\url#1{\texttt{#1}}\fi
\expandafter\ifx\csname urlprefix\endcsname\relax\def\urlprefix{URL }\fi
\providecommand{\bibinfo}[2]{#2}
\providecommand{\eprint}[2][]{\url{#2}}

\bibitem{bednorz:1986}
\bibinfo{author}{Bednorz, J.~G.} \& \bibinfo{author}{Muller, A.}
\newblock \emph{\bibinfo{journal}{Z. Phys. B}} \textbf{\bibinfo{volume}{64}},
  \bibinfo{pages}{189} (\bibinfo{year}{1986}).

\bibitem{wu:1987}
\bibinfo{author}{Wu, M.~K.} \emph{et~al.}
\newblock \bibinfo{title}{Superconductivity at 93 k in a new mixed-phase
  \mbox{Y-Ba-Cu-O} compound system at ambient pressure}.
\newblock \emph{\bibinfo{journal}{Phys. Rev. Lett.}}
  \textbf{\bibinfo{volume}{58}}, \bibinfo{pages}{908} (\bibinfo{year}{1987}).

\bibitem{maeda:1988}
\bibinfo{author}{Maeda, H.}, \bibinfo{author}{Tanaka, Y.},
  \bibinfo{author}{Fukutomi, M.} \& \bibinfo{author}{Asano, T.}
\newblock \bibinfo{title}{A new high-\mbox{T$_c$} oxide superconductor without
  a rare earth element}.
\newblock \emph{\bibinfo{journal}{Jpn. J. Appl. Phys.}}
  \textbf{\bibinfo{volume}{27}}, \bibinfo{pages}{L209} (\bibinfo{year}{1988}).

\bibitem{keimer:2015}
\bibinfo{author}{Keimer, B.}, \bibinfo{author}{Kivelson, S.~A.},
  \bibinfo{author}{Norman, M.~R.}, \bibinfo{author}{Uchida, S.} \&
  \bibinfo{author}{Zaanen, J.}
\newblock \bibinfo{title}{From quantum matter to high-temperature
  superconductivity in copper oxides}.
\newblock \emph{\bibinfo{journal}{Nature}} \textbf{\bibinfo{volume}{518}},
  \bibinfo{pages}{179} (\bibinfo{year}{2015}).

\bibitem{batlogg:2000}
\bibinfo{author}{Batlogg, B.} \& \bibinfo{author}{Varma, C.~M.}
\newblock \bibinfo{title}{The underdoped phase of cuprate superconductors}.
\newblock \emph{\bibinfo{journal}{Phys. World}} \textbf{\bibinfo{volume}{13}},
  \bibinfo{pages}{33} (\bibinfo{year}{2000}).

\bibitem{taillefer:2010}
\bibinfo{author}{Taillefer, L.}
\newblock \bibinfo{title}{Scattering and pairing in cuprate superconductors}.
\newblock \emph{\bibinfo{journal}{Ann. Rev. of Conden. Mat. Phys.}}
  \textbf{\bibinfo{volume}{1}}, \bibinfo{pages}{51} (\bibinfo{year}{2010}).

\bibitem{michon:2023}
\bibinfo{author}{Michon, B.} \emph{et~al.}
\newblock \bibinfo{title}{Reconciling scaling of the optical conductivity of
  cuprate superconductors with planckian resistivity and specific heat}.
\newblock \emph{\bibinfo{journal}{Nat. Comm.}} \textbf{\bibinfo{volume}{14}},
  \bibinfo{pages}{3033} (\bibinfo{year}{2023}).

\bibitem{patel:2019}
\bibinfo{author}{Patel, A.~A.} \& \bibinfo{author}{Sachdev, S.}
\newblock \bibinfo{title}{Theory of a planckian metal}.
\newblock \emph{\bibinfo{journal}{Phys. Rev. Lett.}}
  \textbf{\bibinfo{volume}{123}}, \bibinfo{pages}{066601}
  (\bibinfo{year}{2019}).

\bibitem{carbotte:2011}
\bibinfo{author}{Carbotte, J.~P.}, \bibinfo{author}{Timusk, T.} \&
  \bibinfo{author}{Hwang, J.}
\newblock \bibinfo{title}{Bosons in high-temperature superconductors: an
  experimental survey}.
\newblock \emph{\bibinfo{journal}{Reports on Progress in Physics}}
  \textbf{\bibinfo{volume}{74}}, \bibinfo{pages}{066501}
  (\bibinfo{year}{2011}).

\bibitem{carbotte:1999}
\bibinfo{author}{Carbotte, J.~P.}, \bibinfo{author}{Schachinger, E.} \&
  \bibinfo{author}{Basov, D.~N.}
\newblock \bibinfo{title}{Coupling strength of charge carriers to spin
  fluctuations in high-temperature superconductors}.
\newblock \emph{\bibinfo{journal}{Nature (London)}}
  \textbf{\bibinfo{volume}{401}}, \bibinfo{pages}{354} (\bibinfo{year}{1999}).

\bibitem{schachinger:2000}
\bibinfo{author}{Schachinger, E.} \& \bibinfo{author}{Carbotte, J.~P.}
\newblock \bibinfo{title}{Coupling to spin fluctuations from conductivity
  scattering rates}.
\newblock \emph{\bibinfo{journal}{Phys. Rev. B}} \textbf{\bibinfo{volume}{62}},
  \bibinfo{pages}{9054} (\bibinfo{year}{2000}).

\bibitem{hwang:2006}
\bibinfo{author}{Hwang, J.} \emph{et~al.}
\newblock \bibinfo{title}{\mbox{a}-axis optical conductivity of detwinned
  ortho-\mbox{II} \mbox{YBa$_2$Cu$_3$O$_{6.50}$}}.
\newblock \emph{\bibinfo{journal}{Phys. Rev. B}} \textbf{\bibinfo{volume}{73}},
  \bibinfo{pages}{014508} (\bibinfo{year}{2006}).

\bibitem{hwang:2007}
\bibinfo{author}{Hwang, J.}, \bibinfo{author}{Timusk, T.},
  \bibinfo{author}{Schachinger, E.} \& \bibinfo{author}{Carbotte, J.~P.}
\newblock \bibinfo{title}{Evolution of the bosonic spectral density of the
  high-temperature superconductor \mbox{Bi$_2$Sr$_2$CaCu$_2$O$_{8+\delta}$}}.
\newblock \emph{\bibinfo{journal}{Phys. Rev. B}} \textbf{\bibinfo{volume}{75}},
  \bibinfo{pages}{144508} (\bibinfo{year}{2007}).

\bibitem{heumen:2009}
\bibinfo{author}{van Heumen, E.} \emph{et~al.}
\newblock \bibinfo{title}{Optical determination of the relation between the
  electron-boson coupling function and the critical temperature in
  high-\mbox{$T_c$} cuprates}.
\newblock \emph{\bibinfo{journal}{Phys. Rev. B}} \textbf{\bibinfo{volume}{79}},
  \bibinfo{pages}{184512} (\bibinfo{year}{2009}).

\bibitem{hwang:2021}
\bibinfo{author}{Hwang, J.}
\newblock \bibinfo{title}{Superconducting coherence length of hole-doped
  cuprates from electron–boson spectral density function}.
\newblock \emph{\bibinfo{journal}{Scientific Reports}}
  \textbf{\bibinfo{volume}{11}}, \bibinfo{pages}{11668} (\bibinfo{year}{2021}).

\bibitem{hwang:2015a}
\bibinfo{author}{Hwang, J.}
\newblock \bibinfo{title}{Reverse process of usual optical analysis of
  boson-exchange superconductors: impurity effects on {\it s}- and {\it d}-wave
  superconductors}.
\newblock \emph{\bibinfo{journal}{J. Phys.: Condens. Matter}}
  \textbf{\bibinfo{volume}{27}}, \bibinfo{pages}{085701}
  (\bibinfo{year}{2015}).

\bibitem{shulga:1991}
\bibinfo{author}{Shulga, S.~V.}, \bibinfo{author}{Dolgov, O.~V.} \&
  \bibinfo{author}{Maksimov, E.~G.}
\newblock \bibinfo{title}{Electronic states and optical spectra of \mbox{HTSC}
  with electron-phonon coupling}.
\newblock \emph{\bibinfo{journal}{Physica C}} \textbf{\bibinfo{volume}{178}},
  \bibinfo{pages}{266} (\bibinfo{year}{1991}).

\bibitem{hwang:2004}
\bibinfo{author}{Hwang, J.}, \bibinfo{author}{Timusk, T.} \&
  \bibinfo{author}{Gu, G.~D.}
\newblock \bibinfo{title}{High-transition-temperature superconductivity in the
  absence of the magnetic-resonance mode}.
\newblock \emph{\bibinfo{journal}{Nature (London)}}
  \textbf{\bibinfo{volume}{427}}, \bibinfo{pages}{714} (\bibinfo{year}{2004}).

\bibitem{gotze:1972}
\bibinfo{author}{G\"{o}tze, W.} \& \bibinfo{author}{W\"{o}lfle, P.}
\newblock \bibinfo{title}{Homogeneous dynamical conductivity of simple metals}.
\newblock \emph{\bibinfo{journal}{Phys. Rev. B}} \textbf{\bibinfo{volume}{6}},
  \bibinfo{pages}{1226} (\bibinfo{year}{1972}).

\bibitem{allen:1977}
\bibinfo{author}{Allen, J.~W.} \& \bibinfo{author}{Mikkelsen, J.~C.}
\newblock \bibinfo{title}{Optical properties of \mbox{CrSb}, \mbox{MnSb},
  \mbox{NiSb}, and \mbox{NiAs}}.
\newblock \emph{\bibinfo{journal}{Phys. Rev. B}} \textbf{\bibinfo{volume}{15}},
  \bibinfo{pages}{2952} (\bibinfo{year}{1977}).

\bibitem{puchkov:1996}
\bibinfo{author}{Puchkov, A.~V.}, \bibinfo{author}{Basov, D.~N.} \&
  \bibinfo{author}{Timusk, T.}
\newblock \bibinfo{title}{The pseudogap state in high-\mbox{T$_c$}
  superconductors: an infrared study}.
\newblock \emph{\bibinfo{journal}{J. Phys.: Cond. Matter}}
  \textbf{\bibinfo{volume}{8}}, \bibinfo{pages}{10049} (\bibinfo{year}{1996}).

\bibitem{marel:2003}
\bibinfo{author}{van~der Marel, D.} \emph{et~al.}
\newblock \bibinfo{title}{Quantum critical behaviour in a high-\mbox{T$_c$}
  superconductor}.
\newblock \emph{\bibinfo{journal}{Nature (London)}}
  \textbf{\bibinfo{volume}{425}}, \bibinfo{pages}{271} (\bibinfo{year}{2003}).

\bibitem{berthod:2013}
\bibinfo{author}{Berthod, C.} \emph{et~al.}
\newblock \bibinfo{title}{Non-drude universal scaling laws for the optical
  response of local fermi liquids}.
\newblock \emph{\bibinfo{journal}{Phys. Rev. B}} \textbf{\bibinfo{volume}{87}},
  \bibinfo{pages}{115109} (\bibinfo{year}{2013}).

\bibitem{varma:1989}
\bibinfo{author}{Varma, C.}, \bibinfo{author}{Littlewood, P.},
  \bibinfo{author}{Schmitt-Rink, S.}, \bibinfo{author}{Abrahams, E.} \&
  \bibinfo{author}{Ruckenstein, A.}
\newblock \bibinfo{title}{Phenomenology of the normal state of \mbox{Cu-O}
  high-temperature superconductors}.
\newblock \emph{\bibinfo{journal}{Phys. Rev. Lett.}}
  \textbf{\bibinfo{volume}{63}}, \bibinfo{pages}{1996} (\bibinfo{year}{1989}).

\bibitem{littlewood:1991}
\bibinfo{author}{Littlewood, P.~B.} \& \bibinfo{author}{Varma, C.~M.}
\newblock \bibinfo{title}{Phenomenology of the normal and superconducting
  states of a marginal fermi liquid (invited)}.
\newblock \emph{\bibinfo{journal}{J. Appl. Phys.}}
  \textbf{\bibinfo{volume}{69}}, \bibinfo{pages}{4979} (\bibinfo{year}{1991}).

\bibitem{norman:2006}
\bibinfo{author}{Norman, M.~R.} \& \bibinfo{author}{Chubukov, A.~V.}
\newblock \bibinfo{title}{High-frequency behavior of the infrared conductivity
  of cuprates}.
\newblock \emph{\bibinfo{journal}{Phys. Rev. B}} \textbf{\bibinfo{volume}{73}},
  \bibinfo{pages}{140501(R)} (\bibinfo{year}{2006}).

\bibitem{hwang:2004a}
\bibinfo{author}{Hwang, J.} \emph{et~al.}
\newblock \bibinfo{title}{Marginal fermi liquid analysis of 300 k reflectance
  of \mbox{Bi$_2$Sr$_2$CaCu$_2$O$_{8+\delta}$}}.
\newblock \emph{\bibinfo{journal}{Phys. Rev. B}} \textbf{\bibinfo{volume}{69}},
  \bibinfo{pages}{094520} (\bibinfo{year}{2004}).

\bibitem{bruin:2013}
\bibinfo{author}{Bruin, J. A.~N.}, \bibinfo{author}{Sakai, H.},
  \bibinfo{author}{Perry, R.~S.} \& \bibinfo{author}{Mackenzie, A.~P.}
\newblock \bibinfo{title}{Similarity of scattering rates in metals showing
  t-linear resistivity}.
\newblock \emph{\bibinfo{journal}{Science}} \textbf{\bibinfo{volume}{339}},
  \bibinfo{pages}{804} (\bibinfo{year}{2013}).

\bibitem{legros:2019}
\bibinfo{author}{A. Legros} \emph{et~al.}
\newblock \bibinfo{title}{Universal t-linear resistivity and planckian
  dissipation in overdoped cuprates}.
\newblock \emph{\bibinfo{journal}{Nat. Phys.}} \textbf{\bibinfo{volume}{15}},
  \bibinfo{pages}{142} (\bibinfo{year}{2019}).

\bibitem{sadovskii:2021}
\bibinfo{author}{Sadovskii, M.~V.}
\newblock \bibinfo{title}{Planckian relaxation delusion in metals}.
\newblock \emph{\bibinfo{journal}{Phys.-Usp.}} \textbf{\bibinfo{volume}{64}},
  \bibinfo{pages}{175} (\bibinfo{year}{2021}).

\bibitem{mirzaei:2013}
\bibinfo{author}{Mirzaei, S.~I.} \emph{et~al.}
\newblock \bibinfo{title}{Spectroscopic evidence for fermi liquid-like energy
  and temperature dependence of the relaxation rate in the pseudogap phase of
  the cuprates}.
\newblock \emph{\bibinfo{journal}{PNAS}} \textbf{\bibinfo{volume}{110}},
  \bibinfo{pages}{5774} (\bibinfo{year}{2013}).

\bibitem{hwang:2007a}
\bibinfo{author}{Hwang, J.}, \bibinfo{author}{Timusk, T.} \&
  \bibinfo{author}{Gu, G.~D.}
\newblock \bibinfo{title}{Doping dependent optical properties of
  \mbox{Bi$_2$Sr$_2$CaCu$_2$O$_{8+\delta}$}}.
\newblock \emph{\bibinfo{journal}{J. Phys.: Condens. Matter}}
  \textbf{\bibinfo{volume}{19}}, \bibinfo{pages}{125208}
  (\bibinfo{year}{2007}).

\end{thebibliography}

\newpage

\newpage

\newpage

\newpage

\newpage

\end{document}